\documentclass{zHenriquesLab-StyleBioRxiv}
\usepackage{blindtext}

\leadauthor{Davidson} 

\begin{document}

\title{Helix++: A platform for efficiently \\ securing software}
\shorttitle{Helix++}

\author[1,\Letter]{Jack W. Davidson}
\author[1]{Jason D. Hiser}
\author[1]{Anh Nguyen-Tuong}
\affil[1]{University of Virginia, Department of Computer Science}

\newcommand{\etal}{et.al.\xspace}

\maketitle

\begin{abstract}

The open-source Helix++ project improves the security posture of computing platforms by applying cutting-edge cybersecurity techniques to diversify and harden software automatically. A distinguishing feature of Helix++ is that it does not require source code or build artifacts; it operates directly on software in binary form—even stripped executables and libraries. This feature is key as rebuilding applications from source is a time-consuming and often frustrating process. Diversification breaks the software monoculture and makes attacks harder to execute as information needed for a successful attack will have changed unpredictably. Diversification also forces attackers to customize an attack for each target instead of attackers crafting an exploit that works reliably on all similarly configured targets. Hardening directly targets key attack classes. The combination of diversity and hardening provides defense-in-depth, as well as a moving target defense, to secure the Nation’s cyber infrastructure. 

\end {abstract}



\begin{corrauthor}
\texttt{jwd{@}virginia.edu}
\end{corrauthor}

\section*{Introduction}


Cybersecurity in modern software remains a critical problem that must be addressed for the safety of our personal medical data and devices, critical infrastructure, and banking systems.
Many advances,  including code review techniques, security scanners, fuzzers have helped improve the quality of deployed software.~\cite{li2018fuzzing,bohme2017directed,liang2018fuzzing,zhu2022fuzzing,mcintosh2014impact,mcintosh2016empirical,dos2017investigating,amankwah2020empirical,mburano2018evaluation,roldan2017comparison}  
However, as new bugs inevitably surface, these techniques still require software to be patched and re-deployed. This process can take weeks, months or years!  Redeployment of software is complicated by issues such as compatibility problems which can prevent or delay updates.  Sometimes security patches are ``backported'' to previous versions of software to provide security-enhanced software that is more compatible with existing systems.  Further, even if security updates are ready to be deployed with an update manager, users may delay this process for months due to it requiring downtime (possible in the form of rebooting a machine to install updates.)  Is your Windows/Linux/MacOS machine right now asking to be rebooted to install updates?

End-point detection and prevention techniques also have drawbacks.~\cite{arfeen2021endpoint,yoo2018study}  
They often only detect known threats, and once an attacker knows how to evade the technique, they are of little use.  Further, they are highly prone to false positives.~\cite{ho2012statistical,hu2015false}

Ideally, security could be continuously applied to deployed software.  This approach avoids the cost of re-deployment by applying security features directly to the deployed software.  Also, it avoids the possibility of false positives/negatives during an end-point detection, since the software security is applied to software vetted by system administrators.  Lastly, it allows a pool of randomized software variants to be deployed to make a moving-target defense, thwarting common ``script kiddie'' style attacks, where cyber attacks are automated by exploiting the software monoculture and easily embedding deployed software details into the attack scripts.

Docker containers provides an interesting opportunity to enact this kind of security.  Software systems are designed to be built and re-built upon existing software layers and can efficiently and quickly be re-launched for a user when they need access to software.

Helix++ is our proposed answer to extend docker containers with retrofitted security.  Helix++ leverages modern, robust, transparent, and efficient binary rewriting integrated into a software's Docker build process to create Docker images that can be transparently used by an end user.  The image deployment model is extended to randomly select from a set of equivalent images to avoid deploying the same software to every user, breaking the software monoculture and increasing the attacker's workload.

This paper describes the Helix++ vision, and our progress to date on enabling these possible benefits for wide-scale distribution.  Our current preliminary case study with the University of Virginia ACCORD system indicate that key infrastructure can be protected at modest cost in terms of both analysis time and storage space.


\section*{Helix++}

\begin{figure*}
\centering
\includegraphics[width=\linewidth,clip,trim=1.0in 2.85in 1.0in 2.0in]{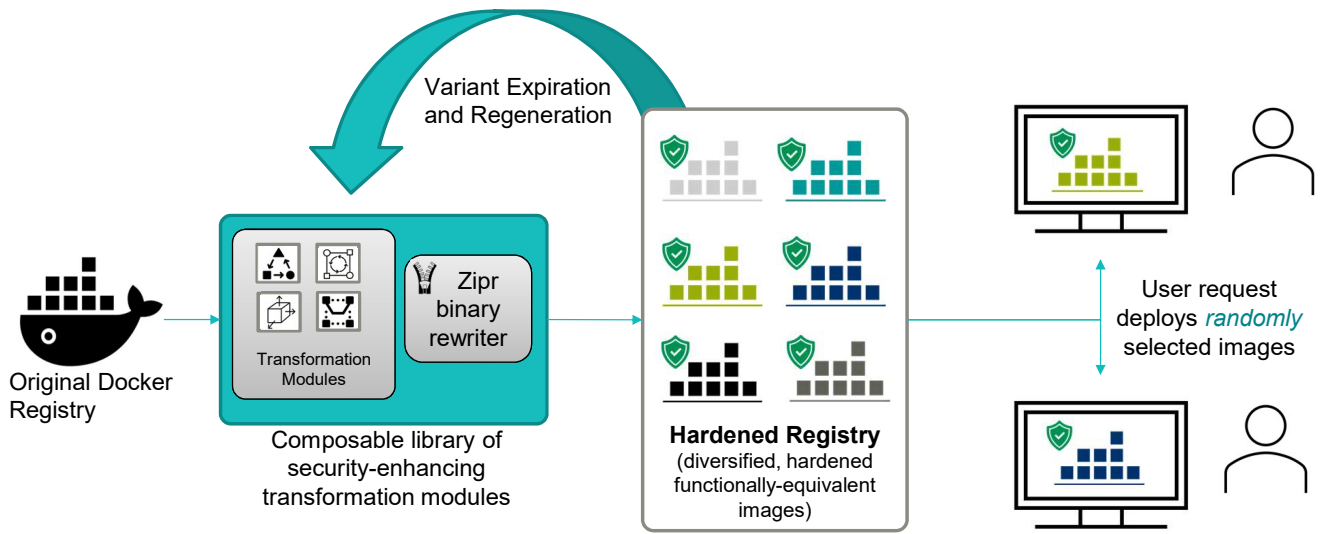}
\caption{The Helix++ architecture.}
\label{fig:helix++}
\end{figure*}

Figure~\ref{fig:helix++} shows the overall view of the Helix++ architecture.  
The key idea is to take a repository full of software ready to deploy (e.g., a Docker Registry), apply security hardening and diversity techniques to the software in that registry, and build a hardened registry full of functionally equivalent software that has been hardened and diversified.   

Each piece of software is in the hardened registry multiple times with a different random seed used for the pseudo-random diversification transforms.  
This feature allows each user request for software to get a diverse variant of the software.  
In theory, every user request could get a piece of software that has never been deployed before.  However, in practice the registry may not be large enough or the transformer fast enough to maintain unique software for each request during periods of high demand.   

Our research goals are to understand request frequency distributions, computational cost for variant generation, and the system costs and benefits.

\paragraph{Binary Rewriting.} Helix++ is based on binary rewriting.
Binary rewriting allows modification of the behavior of an executable file without access to its source code. This feature is particularly useful for security researchers who need to analyze and modify the behavior of a binary file. With binary rewriting users can modify the code of an executable file by replacing or inserting instructions, changing function parameters, and manipulating data structures. This capability allows users to add security checks or patch vulnerabilities in software applications.

Many binary rewriters have an API for building plugins.  A plugin can typically  transform a program and plugins can be composed to combine functionality.  For example, one plugin might transform the stack layout, while another plugin might rearrange the layout of global variables.  Together, both stack and global variable locations can be randomized.

Helix++ leverages the Zipr static binary rewriter, which is described more in Helix++ State of Development Section.

\paragraph{Hardened Registry.}
A registry full of variants of security-hardened programs can offer several benefits to users and organizations. Firstly, it provides a centralized location for accessing a range of security-hardened programs, which reduces the time and effort needed to locate, download, and install secure software. This approach can be particularly beneficial for organizations that need to ensure the security of their systems and data, as it allows them to easily access and deploy secure software across their network. 

Another benefit of a registry full of security-hardened programs is that it can help promote security best practices and standards. By providing users with access to secure software, it encourages the use of security-conscious practices, such as keeping software up-to-date. Furthermore, having a registry of security-hardened programs can help promote collaboration and knowledge sharing between security professionals, as they can contribute their own security transformations to Zipr, improving the security of the registry. This benefit can help to improve the overall security of the software ecosystem, benefiting all users who rely on these programs to keep their systems and data secure.

A key feature of the registry is that variants can expire. With sufficient compute resources, it may be possible to expire a variant after it is deployed just one time.

\section*{Helix++ State of Development}
\label{sec:state}
Helix++ is being built from several stable industrial software components and the Zipr static binary rewriter.

\paragraph{Binary Rewriting with Zipr.}

Helix++'s binary rewriting is based on Zipr.~\cite{hawkins2017zipr,hawkins2017securing,hiser2017zipr++,zipr_repo,zipr_cookbook,zipr_sdk}
Zipr's core functionality supports block-level instruction randomization (BILR), similar to Zhan, \etal~\cite{zhan2014defending}  Zipr achieves this functionality by doing deep binary analysis and building an IR.  Zipr's IR includes every instruction in the program, the static data object in the program (globals, switch tables, ELF linking tables, etc.), exception handling information, and meta-data about the program (the target architecture bit width, etc.)  Zipr's IR is similar to a low-level compiler's back-end IR.  After building the IR, Zipr can invoke Zipr plugins built against the Zipr API.  The API allows for easy composition of plugins, but of course the plugins have to be robust enough to work together.  For example, if one plugin converts indirect branches to direct branches with an \texttt{if/then/else} construct, and a subsequent plugin  instruments indirect branch instructions with a control flow integrity technique, the subsequent plugin will not find any indirect branches to instrument and the two plugins would not compose well.  However, if one plugin instruments stack operations and another plugin instruments global data operations, the two should compose without any special consideration.  The Zipr plugin API allows typical, low-level modifications to the IR -- insert, modify or delete any portion of the IR.  Further, it allows the user to selectively re-run the deep static analysis techniques used to build the initial IR on the modified IR. This feature can be useful, for example, when instrumentation uses a dead register for performance optimizations, and later optimizations need to know which registers are still dead.

These features make Zipr one of the most stable, efficient and effective static binary rewriters resulting from millions of dollars of funding.  
First commits to the project's source code are from the year 2010.  
Zipr transforms binary programs, stripped or not, and generates a functionally equivalent binary program.  
It is most robust on x86-64, Linux ELF binaries, but also has support for ARM ELF binaries, MIPS ELF Binaries and Windows PE/PE+ files.   For all file formats, both 32- and 64-bit architectures are supported.

Zipr has been demonstrated to be one of very few robust binary rewriters~\cite{schulte2022binary}.  Zipr commits are regularly tested against 42 real world software applications compiled with a variety of compilers (gcc, clang, ollvm, icx), compiler flags (O0, O1, O2, O3, Ofast, OSize), in both PIE and non-PIE mode, in both stripped and unstripped form, and across 3 different flavors of Ubuntu (16.04 LTS, 18.04 LTS, and 22.04 LTS).  The selected programs are comprised of C, C++ and Rust applications.

Zipr has been used for a variety of projects, including the DARPA Cyber Grand Challenge (CGC).  CGC is an autonomous capture the flag contest where cyber reasoning systems played the game.  Zipr was part of the TechX team's submission, Xandra.  Zipr was used to generate the best security score of all performers and placed 2nd overall.~\cite{cgc,nguyen2018xandra}  

Zipr was also used in DARPA's Cyber Fault-tolerant Attack Recovery (CFAR) Program~\cite{cfar}.  The program's goal was to generate diverse variants of web servers, and run these programs in parallel.  If the variants diverged in behavior, one could assume that a cyber attack was occurring and remediative actions could be taken.  Zipr was used to generate variants with varying code, stack, global, and heap layout, provably preventing exploits against several important classes of common memory errors.  Because Zipr's technology is agnostic to the source language, it was the only solution in the program that was able to handle the ADA Web Server application (obviously written in ADA.)

in addition these projects, Zipr has been used to do antifragility work, and is the basis for effective binary-only fuzzing with tools like  Zafl and the binary-only based version of Untracer, called HeXcite.~\cite{leach2022start,zafl_source,nagy2021same,nagy2021breaking,nagy2019full}

\paragraph{Hardened Registries in UVA ACCORD.}

As a practical example of how to use hardened registries in a real-world environment, we have partnered with the ACCORD team at the University of Virgnia (UVA).  
ACCORD is: 
\begin{quote}
     a web-based platform which allows researchers from public universities across the state of Virginia to analyze and store their sensitive data in a central location.

ACCORD is appropriate for de-identified PII, FERPA, de-identified HIPAA, business confidential, and other types of de-identified sensitive data

Thanks to funding provided by the National Science Foundation (Award \#: 1919667), ACCORD is available at no cost to researchers in the state of Virginia.~\cite{accord_description}
\end{quote}

To copy data to/from the ACCORD secure storage, one needs to use Globus.~\cite{globus}  To examine or compute on this data, one needs to run a ``session''.  The user interacts with a web front end to start these sessions.  Current sessions include  a C/C++ IDE, a Python interpreter, or R studio.~\cite{theia,rstudio}

Each session is deployed in Kubernetes pod.~\cite{kubernetes}.  Each pod contains two docker containers, one that the user directly interacts with, and a second side-car container that mediates all network traffic to/from the user-controlled container for security.  Current security policies allow the user to use \texttt{pip} to install python packages from a set of ``trusted'' repositories.  While the repositories are generally thought to be secure, one is certain about the provenance or security that's generally in the python repositories.  Thus, python is a key vector that the administrators worry about with regard to security.

In light of this concern, we have worked with the ACCORD team to apply basic Zipr protections to the python interpreter.  So far, we have limited ourselves to this key piece of software.  We have created multiple versions of the ACCORD Docker images.  Creating these images is done by extending the Github workflows with new jobs that alter the images created in previous steps.  Leveraging the massive parallelism provided by Github, only minutes of additional processing time is added to the workflows.

We have further  made a slight modification to ACCORD's web interface to randomly select from the set of functionally equivalent containers when starting a new session.  While these mechanisms are fully functional to the best of our knowledge, we are working with the ACCORD team to gradually deploy these changes to the end users of the ACCORD system.  At the current time, we need to discuss the details of this procedure with the ACCORD administrators, but they are excited about the general plan to add security to their system.

If successful, hardened containers will be an end goal of the Helix++ project.  We plan to measure features such as increased storage space for the registry, time to transform images, and the what replacement rate are manageable by our infrastructure.


\section*{Conclusions}

Helix++ is a system where a binary rewriter is used to add security hardening and diversity transformations in deployed software.  A docker registry full of hardened variants is created.  These variants are used to satisfy user requests for software.  An expiration policy ensures that variants are always fresh, such that a software monoculture never occurs.  Further, when additional hardening is available for the software, they can be automatically re-applyed with the binary rewriter.  Leveraging the Zipr binary rewriter and its suite of transforms, we have built upon the UVA ACCORD system to randomly select variants for use by end users.  While the project is still ongoing, we have made excellent progress and the ACCORD staff are excited for the opportunity to secure their system further.

\begin{acknowledgements}

This material is based upon work supported by the National Science Foundation under Grant No. 2115130.  Any opinions, findings, and conclusions or recommendations expressed in this material are those of the author(s) and do not necessarily reflect the views of the National Science Foundation.

\end{acknowledgements}

\section*{Bibliography}

\bibliography{citations}

\end{document}